\documentclass[sigconf]{acmart}
\usepackage{multicol}
\usepackage{enumitem}
\AtBeginDocument{%
  \providecommand\BibTeX{{%
    \normalfont B\kern-0.5em{\scshape i\kern-0.25em b}\kern-0.8em\TeX}}}

\setcopyright{acmcopyright}
\copyrightyear{2018}
\acmYear{2018}
\acmDOI{10.1145/1122445.1122456}

\acmConference[Woodstock '18]{Woodstock '18: ACM Symposium on Neural
  Gaze Detection}{June 03--05, 2018}{Woodstock, NY}
\acmBooktitle{Woodstock '18: ACM Symposium on Neural Gaze Detection,
  June 03--05, 2018, Woodstock, NY}
\acmPrice{15.00}
\acmISBN{978-1-4503-XXXX-X/18/06}

\copyrightyear{2022} 
\acmYear{2022} 
\setcopyright{acmlicensed}\acmConference[DDAM '22]{Proceedings of the 1st International Workshop on Deepfake Detection for Audio Multimedia}{October 14, 2022}{Lisboa, Portugal}
\acmBooktitle{Proceedings of the 1st International Workshop on Deepfake Detection for Audio Multimedia (DDAM '22), October 14, 2022, Lisboa, Portugal}
\acmPrice{15.00}
\acmDOI{10.1145/3552466.3556530}
\acmISBN{978-1-4503-9496-3/22/10}

\begin{document}

\title{Fully Automated End-to-End Fake Audio Detection}



\author{Chenglong Wang}
\affiliation{%
  \institution{University of Science and Technology of China, Hefei, China}
  \institution{NLPR, Institute of Automation, Chinese Academy of Sciences}
  \city{Beijing}
  \country{China}
}
\email{chenglong.wang@nlpr.ia.ac.cn}

\author{Jiangyan Yi}
\affiliation{%
  \institution{NLPR, Institute of Automation, Chinese Academy of Sciences, China}
  \institution{School of Artificial Intelligence, University of Chinese Academy of Sciences}
  \city{Beijing}
  \country{China}}
\email{jiangyan.yi@nlpr.ia.ac.cn}

\author{Jianhua Tao}
\affiliation{%
	\institution{NLPR, Institute of Automation, Chinese Academy of Sciences, China}
	\institution{School of Artificial Intelligence, University of Chinese Academy of Sciences, China}
	\institution{CAS Center for Excellence in Brain Science and Intelligence Technology}
	\city{Beijing}
	\country{China}}

\author{Haiyang Sun}
\affiliation{%
	\institution{NLPR, Institute of Automation, Chinese Academy of Sciences, China}
	\institution{School of Artificial Intelligence, University of Chinese Academy of Sciences}
	\city{Beijing}
	\country{China}}

\author{Xun Chen}
\affiliation{%
	\institution{University of Science and Technology of China}
	\city{Hefei}
	\country{China}}

\author{Zhengkun Tian}
\affiliation{%
	\institution{NLPR, Institute of Automation, Chinese Academy of Sciences, China}
	\institution{School of Artificial Intelligence, University of Chinese Academy of Sciences}
	\city{Beijing}
	\country{China}}

\author{Haoxin Ma}
\affiliation{%
	\institution{NLPR, Institute of Automation, Chinese Academy of Sciences, China}
	\institution{School of Artificial Intelligence, University of Chinese Academy of Sciences}
	\city{Beijing}
	\country{China}}

\author{Cunhang Fan}
\affiliation{%
	\institution{School of Computer Science and Technology, Anhui University}
	\city{Hefei}
	\country{China}}

\author{Ruibo Fu}
\affiliation{%
	\institution{NLPR, Institute of Automation, Chinese Academy of Sciences, China}
	\institution{School of Artificial Intelligence, University of Chinese Academy of Sciences}
	\city{Beijing}
	\country{China}}

\renewcommand{\shortauthors}{Trovato and Tobin, et al.}

\begin{abstract}
  The existing fake audio detection systems often rely on expert experience to design the acoustic features or manually design the hyperparameters of the network structure. However, artificial adjustment of the parameters can have a relatively obvious influence on the results. It is almost impossible to manually set the best set of parameters. Therefore this paper proposes a fully automated end-to-end fake audio detection method. We first use wav2vec pre-trained model to obtain a high-level representation of the speech. Furthermore, for the network structure, we use a modified version of the differentiable architecture search (DARTS) named light-DARTS. It learns deep speech representations while automatically learning and optimizing complex neural structures consisting of convolutional operations and residual blocks. The experimental results on the ASVspoof 2019 LA dataset show that our proposed system achieves an equal error rate (EER) of 1.08\%, which outperforms the state-of-the-art single system.
\end{abstract}




\maketitle

\section{Introduction}
In the last few years, the development of speech synthesis technology \cite{wang2020bi, wang2017tacotron, sotelo2017char2wav, valin2019lpcnet} is increasing rapidly driven by deep learning. These models can generate high quality audio that is comparable to the human voice. Although the technology itself has no negative attributes, it is easy to be misused, such as faking audio to commit fraud and spread public opinion. Therefore, an increasing number of work \cite{yi2021half, ma2021continual, yamagishi2021asvspoof, todisco2019asvspoof, kinnunen2017asvspoof, yi2022add} have been focused on detecting the fake audio recently.

In order to improve the performance of fake audio detection systems, recent works have focused on two aspects: improving the acoustic features of audio and designing effective classification models. It is particularly important to select acoustic features that can effectively distinguish fake audios from real audios. Todisco et al. \cite{todisco2017constant} apply the constant Q-transform instead of the short-time Fourier transform to process speech signals, which outperforms the Mel frequency cepstrum coefficients(MFCC). Sahidullah et al.  \cite{sahidullah2015comparison} replace the Mel scale filters with linear filters and propose linear frequency cepstrum coefficients(LFCC), making it more focused on high frequency band features compared to MFCC. With the development of unsupervised pre-training, Yang et al. \cite{xie2021siamese} first use the wav2vec pre-trained model as the feature extractor which obtains more robust acoustic feature. At the latest ADD2022 challenge \cite{yi2022add}, Donas et al. \cite{martin2022vicomtech} also propose to use an improved version of wav2vec 2.0 as a feature extractor and win the first place on Track 1. The input to wav2vec feature extractor is the raw waveform, which learns speech information from a large amount of unlabeled speech that may contain information useful for fake audio detection.

Another approach is to design an effective classification model that learns a distinguished representation of real and fake audio. Gaussian mixture model (GMM) is the traditional classification model. With the development of deep learning, the performance of convolutional neural networks (CNN) \cite{lai2019assert, alzantot2019deep} is better than GMM. For example, light convolution neural network (LCNN) \cite{lavrentyeva2017audio} with max feature map (MFM) activation function can not only separate noise signals from information signals by competitive learning, but also play a role in feature selection. The residual network (ResNet) \cite{he2016deep} proposes a residual module that addresses the problem of network degradation. Compared with hand-designed approaches, differentiable architecture search (DARTS) \cite{liu2018darts} makes great achievements in the design of deep neural networks to automate the manual process of architecture design. Ge et al. \cite{ge2021partially} first successfully use a Partially-Connected DARTS approach to the fake audio detection tasks. Although Ge et al. \cite{ge2021partially} implement automation in the network structure, they choose LFCC and fast Fourier transform (FFT) in the feature level, which still requires manual parameter setting. For example, LFCC features need to set the window length, fft transform length, filter set coefficients and so on. However, artificial adjustment of the parameters can have a relatively obvious influence on the results. It is almost impossible to manually set the best set of parameters. Therefore, it’s necessary to design a fully automated fake audio detection system. 

To address these problems, we propose a fully automated end-to-end fake audio detection. More generally, we use pre-trained wav2vec model as feature extractor instead of traditional acoustic features. Inspired by light-CNN \cite{wu2018light}, we propose a light-DARTS based on DARTS to learn deep speech representations automatically. As shown in Figure 1 (d), we apply the max feature map (MFM) module to DARTS, which plays the role of feature selection. The main contributions of this study can be summarized as follows:

\begin{itemize}
	\item To our best knowledge, this is the first work to propose light-DARTS for fake audio detection task.
	\item We propose a a fully automated end-to-end fake audio detection method. We use wav2vec features as inputs for ligth-DARTS.
	
\end{itemize}

The experimental results on the ASVspoof 2019 LA dataset show that our proposed method can acquire the EER of 1.08\%. This result demonstrates the effectiveness of our proposed method.

The rest of this paper is organized as follows: Section 2 presents the related work. Section 3 illustrates the our proposed method. Experiments, results and discussions are reported in Section 4 and 5, respectively. Finally, the paper is concluded in Section 6.

\begin{figure*}[t]
	\centering
	
	\includegraphics[height=9.4cm,width=0.95\textwidth]{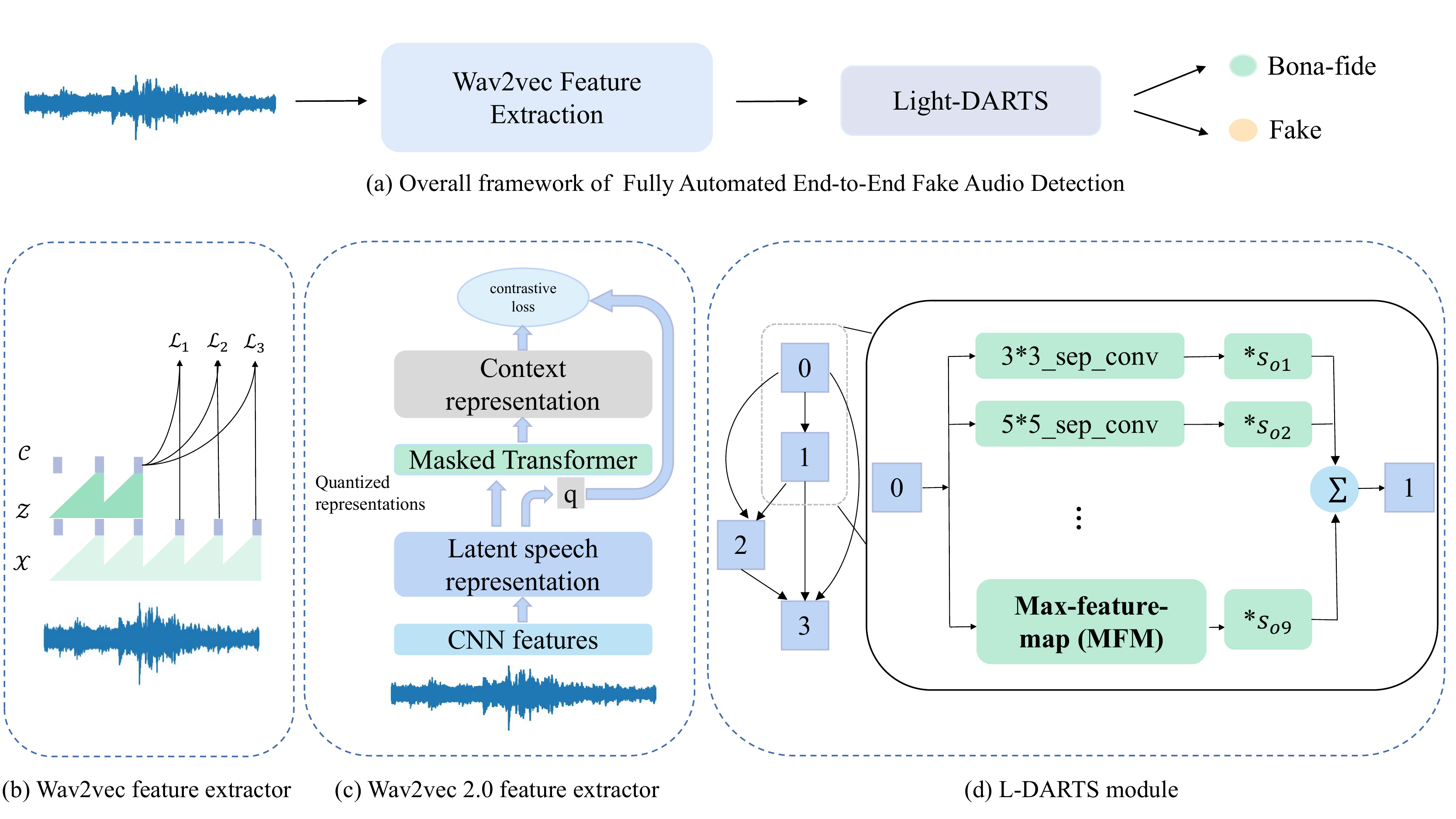}
	\caption{
		(a) Overall framework of the fully automated fake audio detection system. The system consists of the feature extraction module and the light-DARTS module.
		(b) Illustration of the wav2vec feature exctraction module. 
		(c) Illustration of the wav2vec 2.0 feature exctraction module.
		(d) Illustration of the light-DARTS module. We apply the max feature map (MFM) module to DARTS.
	}
	
	\label{fig:speech_production}
\end{figure*}

\section{Related Work}
\subsection{DARTS}

 Compared with hand-designed approaches, DARTS [3] has made great achievements in the design of deep neural networks to automate the manual process of architecture design. Liu et al. \cite{liu2018darts} proposed the DARTS algorithm to use softmax to serialize the discrete search space for gradient-based structural search of neural networks. Instead of fixing the execution of a single operation at a specific layer, the authors define the structure search as a convex optimization problem by selecting the optimal operation from a set of operations and then alternating the operational parameters of the network and the structural parameters corresponding to the operations by gradient descent. Specifically, the learnable parameters of the operations in the network are optimized on the training dataset and the structural parameters corresponding to the different operations are trained on the validation dataset. Ultimately, a sparse network structure is obtained by solving this convex optimization problem to select the most appropriate operation for each layer.

\subsection{Self-supervised Pretraining Models}
Self-supervised pre-training becomes a hot issue in recent years and has been applied in the field of natural language processing and computer vision already. For speech processing, more and more self-supervised speech models are proposed, such as wav2vec \cite{baevski2020wav2vec, schneider2019wav2vec}, HuBERT \cite{hsu2021hubert}, wav2vec 2.0 which allowing to learn high-level representations of the speech signal. The above methods have been successfully applied in the fields of speech recognition, speaker recognition and emotion recognition \cite{fan2020exploring, pepino2021emotion}. As for fake audio detection, only a few works have explored this similar approach \cite{lv2022fake, wang2021investigating, martin2022vicomtech}. Motivated by \cite{lv2022fake}, we propose to use a pre-trained wav2vec style model as a feature extractor instead of traditional acoustic features, which solves exactly the problem of acoustic features requiring manual parameter setting. Furthermore, the features extracted from models pretrained on large-scale datasets help reduce overfitting, thereby improving reliability and domain robustness. 

\section{Our Proposed System}

As shown in Figure 1 (a), the fully automated end-to-end fake audio detection consists of two modules: (1) The feature extraction module: It consists of a wav2vec feature extractor, which obtains a high-level representation of the speech. (2) The Light-DARTS module: We apply the max feature map (MFM) module to DARTS, which plays the role of feature selection. 

\subsection{Wav2vec Feature Extractor}

The input to wav2vec \cite{schneider2019wav2vec} feature extractor is the raw waveform, which learns speech information from a large amount of unlabeled speech that may contain information useful for fake audio detection. In addition, pre-trained models trained from a large amount of unlabeled information can yield more generalized generic speech features, which may also be useful for distinguishing real from fake speech. Therefore, wav2vec is used as a feature extractor in this paper.

\subsubsection{Pretrained wav2vec}

 As shown in Figure 1 (b), the wav2vec model contains two convolutional neural networks, an encoder network that maps the original input audio signal to a hidden space, and a context network that combines multiple time step outputs of the encoder network. The encoder generates a representation $z_i$ for each time step $i$, while the context network combines the multiple encoder time step outputs into a new representation $c_i$ for each time step $i$. 

\subsubsection{Pretrained wav2vec 2.0}
As shown in Figure 1 (c), the wav2vec 2.0 \cite{baevski2020wav2vec} model includes three stages. Firstly, the raw speech is send into a local encoder which contains several convolutional layers (CNN). The stride of the local encoder is 20ms and the receptive field is 25ms. The local encoder embeddings are then fed as input to the context encoder, which consists of several transformer modules. In the open source two pre-trained models, the base model contains 12 transformer blocks with 8 attention heads per block, while the large model contains 24 transformer blocks with 16 attention heads per block. At the third stage, the local encoder representation is input to the quantization module, which consists of 2 codebooks, each with 320 entries. The linear mapping then converts the local encoder representation into logits. Given the logits, Gumbel-Softmax \cite{jang2016categorical} is applied to a sample of each codebook. The selected codes are concatenated and the resulting vectors are linearly transformed to obtain a quantized representation of the local encoder output, which is used in the objective function. The high-level embeddings is represented as: 
\begin{equation}
	\hat{X}_{embedding}(t) = W2V(X(t)) \in \mathbb{R}^{T*F}
\end{equation}
where $ W2V(\cdot) $ denotes the pretrained wav2vec model or wav2vec2.0 model, which aims to obtain high-level representations $\hat{X}_{embedding}(t)$ of raw speech $ X(t) $.

\subsection{Light-DARTS}
Inspired by light-CNN \cite{wu2018light}, we propose a light-DARTS based on DARTS to learn representations of wav2vec features automatically. Search space, using a cell-based approach for spatial definition, finds a computational cell as the building block of the final architecture, and the learned cells can be stacked to form a convolutional network. By training out one small cell to form a large network. Suppose the node is represented as $x^{(i)}$ , then each directed edge $(i, j)$ will be associated with a node $x^{(i)}$ with some operation $o^{(i,j)}$ associated with it. To construct a continuous search space and generate an architecture, the predecessors and edges must be computed, i.e., a continuous relaxation optimization must be performed. First, the intermediate nodes based on the predecessor nodes are calculated as follows.
\begin{equation}
	x^{(j)} = \sum_{i<j}o^{(i,j)}(x^{(i)}) 
\end{equation}

where $o$ denotes a set of candidate operations (e.g., convolution, max-pool, zero), and each operation denotes a function $o(\cdot)$ to be applied to the node $x^{(i)}$.

Secondly, in order to make the search space continuous, we relax the classification choice of a particular operation to $softmax$, among all possible operations, by the following formula.
\begin{equation}
	\overline{o}^{(i,j)}(x) = \sum_{o \in \mathcal{O}}\frac{\exp(\alpha_o^{(i,j)})}{\sum_{o^\prime \in \mathcal{O}}\exp(\alpha_{o^\prime} ^{(i,j)})} o(x)
\end{equation}

where the operational mixing weights of a pair of nodes $(i, j)$ are parameterized by a vector of dimension $\left|O\right|$ parametrized. The task of architectural search will be reduced to learning a set of continuous variables $\alpha = {\alpha^{(i,j)}}$. Finally replacing each mixed operation with the most probable operation $\overline{o}^{(i,j)}$, to obtain a discrete architecture. The formula is as follows.
\begin{equation}
	o^{(i, j)} = \arg \max_{o \in \mathcal{O}} \alpha_o^{(i,j)}
\end{equation}

Then, light-DARTS jointly learns operation parameters $\alpha$ and weight parameters $w$ (e.g., convolution kernels) using the gradient descent method by solving a bilevel optimization problem, where operation parameters $\alpha$ are upper-level variables and weight parameters $w$ are lower-level variables:
\begin{equation}
	\begin{split}
		\min_\alpha \mathcal{L}_{val}(\omega^*(\alpha),\alpha) \\ 
		\omega^*(\alpha) = {\arg\min}_\omega \mathcal{L}_{train}(\omega,\alpha)
	\end{split}
\end{equation}
$\mathcal{L}_{train}$ and $\mathcal{L}_{val}$  denote the training and validation losses.

The key to handling edges is the combination of node connections and activation functions into a matrix, where each element represents the weight of the connection and activation function. When searching, light-DARTS traverses all nodes and finds the average weight of the relevant connections, thus turning the search space into a continuous space. The objective function becomes differentiable, and both the structure weights and the network weights are optimized. At the end of the search, the connection and activation function with the highest weight are selected to form the final network.

Furthermore, we introduce the MFM module in the architecture search space, which plays the role of feature selection. The MFM operation which combines two feature maps and outputs element-wise maximum one as shown in Fig. 2 can be written as:
\begin{equation}
	h(x) = \max{(x_1,x_2)}
\end{equation}

By using MFM, we select and retain 50\% of the information from the input feature maps through element-level maximum operations across feature channels. 

We define all possible candidate operations in a neural cell in the search space $\mathcal{O}$, containing the following nine operations.

\setlength\columnsep{0.01cm}
\begin{multicols}{2}
	\begin{itemize}[leftmargin=0.5cm]
		
		\item 3*3 separable convolutions 
		\item 5*5 separable convolutions
		\item 3*3 dilated convolutions
		\item 5*5 dilated convolutions
		\item 3*3 average pooling
		\item 3*3 max pooling
		\item skip connnection
		\item zero
		\item max feature map
		
	\end{itemize}	
\end{multicols}

\section{Experiments}

\subsection{Dataset}
\subsubsection{ASVspoof 2019 challenge database}

Our focus is on the interference of speech synthesis and speech conversion with real speech, so we use the ASVspoof2019 logical access (LA) dataset \cite{todisco2019asvspoof}. The LA data set contains three subsets: training set, development set and evaluation set. 

\subsubsection{ASVspoof 2021 challenge database}

We focused on the deep fake (DF) tracks for the ASVSpoof2021 challenge \cite{yamagishi2021asvspoof}, where the training and development sets are the same as the ASVspoof 2019 LA database, but the evaluation set differs. The DF evaluation set shows audio coding and compression artifacts which includes approximately 600K of audio processing with various commercial audio codecs.

\subsubsection{ADD 2022 challenge database}

We focused on the low-quality fake audio detection(LF) track for the ADD2022 challenge. The train and development sets contain clean speech based on the multi-speaker Mandarin speech corpus AISHELL-3 [19]. The training set contains 27084 audio and the development set contains 28324 audio. The evaluation set contains a variety of ambient noise and background music.

\subsection{Evaluation Metrics}

In this work, in order to evaluate the results of different fake audio detection systems, the equal error rate (EER) is used as the evaluation metrics. The EER is the operating point where the false rejection rate (FRR) and false acceptance rate (FAR) are equal.

\subsection{Experimental Setup}

The wav2vec pretrained model variant “wav2vec large”, which we use as a pretrained feature extractor using additional linear transformations and a larger context network, is trained on 960 hours Librispeech \cite{panayotov2015librispeech}. The model’s downsampling factor is 160. Thus, there is a 512-dimensional vector for every 10 ms of speech. To form batches, the 400 time frames are fixed by truncating or concatenating. Therefore, the shape of wav2vec feature is $400*512$. For wav2vec 2.0 feature extractor, we have adopted two versions: ‘wav2vec 2.0 base’ and ‘wav2vec 2.0 large’. The base model contains 768 model dimension, 12 transformer blocks, inner dimension (FFN) 3,072 and 8 attention heads. The large model contains 24 transformer blocks with model dimension 1,024, inner dimension 4,096 and 16 attention heads. Like the wav2vec features, we also performed truncating or concatenating in the temporal dimension. Therefore, the shape of wav2vec 2.0 base and large are $400*768$ and $400*1024$, respectively.

In the search stage, we followed the same experimental setup as the original DARTS \cite{liu2018darts} study. The neural cells were divided into two categories, namely normal cells and reduction cells. The number of cells N is 8, contains 6 normal cells and 2 reduction cells. The reduction cells were inserted into the 1/3 and 2/3 locations of the entire network. At each step, the model is trained for 50 epochs by Adam optimiser with the learning rate 0.0001.

\section{Results and Discussion}

\begin{table}[t]
	
	\caption{ The results of EER(\%) for our proposed different systems on ASVspoof2019 LA.}
	
	\centering
	\begin{tabular}{cccc }
		\toprule
		\textbf{Feature} & \textbf{Network Architecture} & \textbf{Dev} & \textbf{Eval} \\
		\midrule
		LFCC & LCNN &  0.13 & 4.75         \\
		Wav2vec   & LCNN  & 0.03  & 3.51         \\
		Wav2vec 2.0-base  & LCNN &  0.02  &  3.32      \\
		Wav2vec 2.0-large & LCNN &  0.14  &  3.56       \\
		LFCC  &  DARTS  & 0.01 &    4.82         \\
		Wav2vec & DARTS & 0.02 &  2.18     \\
		Wav2vec 2.0-base & DARTS & 0.01 & 2.23 \\
		Wav2vec 2.0-large & DARTS & 0.05 & 1.97  \\
		LFCC  &  light-DARTS  & 0.05 &   4.35          \\
		Wav2vec & light-DARTS & 0.06 &  1.51     \\
		Wav2vec 2.0-base & light-DARTS & 0.01 & 1.19  \\
		Wav2vec 2.0-large & light-DARTS & 0.02 & \textbf{1.08}  \\
		
		\bottomrule
	\end{tabular}
	
\end{table}

\begin{table}[t]
	
	\caption{The results of EER(\%) for our proposed different systems on ASVspoof2021 DF. For the characteristics of this test set, we also evaluated narrow-band (NB) FIR filters using a data enhancement approach, following a similar procedure as in [21]. Where T23 refers to the first place team of the ASVspoof2021 DF track.}
	
	\centering
	\begin{tabular}{ccc }
		\toprule
		\textbf{Feature} & \textbf{Network Architecture} & \textbf{DF} \\
		\midrule
		LFCC  &  light-DARTS  &   16.34  \\
		Wav2vec & light-DARTS &    9.21   \\
		Wav2vec 2.0-base & light-DARTS & 8.16  \\
		Wav2vec 2.0-large & light-DARTS &\textbf{7.86}  \\
		T23  &  best result in ASVspoof2021 \cite{yamagishi2021asvspoof} &   15.64  \\
		\bottomrule
	\end{tabular}
	
\end{table}

\begin{table}[t]
	
	\caption{The results of EER(\%) for our proposed different systems on ADD2022 Track 1. For the characteristics of this test set, we also evaluated narrow-band (NB) FIR filters using a data enhancement approach, following a similar procedure as in [21]. Where A01 refers to the first place team of the ADD2021 Track 1.}
	
	\centering
	\begin{tabular}{ccc }
		\toprule
		\textbf{Feature} & \textbf{Network Architecture} & \textbf{Track 1} \\
		\midrule
		LFCC  &  light-DARTS  &   24.18  \\
		Wav2vec & light-DARTS &    22.52   \\
		Wav2vec 2.0-base & light-DARTS & 21.23  \\
		Wav2vec 2.0-large & light-DARTS &\textbf{20.11}  \\
		A01 &  best result in ADD 2022 \cite{yi2022add} &   21.71  \\
		\bottomrule
	\end{tabular}
	
\end{table}

\begin{table}[t]
	
	\caption{Performance comparison between the proposed model and other models on the evaluation set of ASVspoof2019 LA.}
	
	\centering
	\begin{tabular}{ccc }
		\toprule
		\textbf{Metheds} & \textbf{Dev} & \textbf{Eval} \\
		\midrule
		W2V-XLSR-LGF \cite{wang2021investigating}  & - &    1.28         \\
		FFT-L-SENet \cite{zhang122021effect}  & -  &  1.14          \\
		Attention-based CNN \cite{ling2021attention}     & 0.16 &  1.87         \\
		RAWNET2 \cite{tak2021end}   & -   &  1.12         \\
		LFCC-PC-DARTS \cite{ge2021partially}  & 0.002  & 4.87           \\
		W2V-Siamese \cite{xie2021siamese}    & 0.004 & 1.15     \\
		W2V2.0-light-DARTS(ours)    & 0.02 & \textbf{1.08} \\
		
		\bottomrule
	\end{tabular}
	
\end{table}

\subsection{Results on ASVspoof2019}
Table 1 shows the results of our proposed different systems. We can find the following observations. Firstly, the wav2vec features perform better than the LFCC features, one of the accepted baselines. More specifically, the EER of w2v features is 3.51\% while the EER of LFCC features is 4.75\% with both fixed LCNN as back-end classifiers. The reason is that the features extracted from models pre-trained on large-scale datasets help reduce overfitting, thereby improving reliability and domain robustness. In addition, when the pre-trained model is replaced with wav2vec 2.0-base, the result is further improved to 3.32\%. A possible explanation for this might be that compared to wav2vec, wav2vec 2.0 employs the transformer architectures to model the dependencies between feature sequences, resulting in a more discriminative representation of speech. Interestingly, the EER of wav2vec 2.0-large is observed to 3.56\%, even slightly higher than wav2vec. This inconsistency may be due to the fact that the speech representation obtained from the pre-trained model of wav2vec 2.0-large is not adapted to the LCNN structure. These results indicate that the wav2vec pre-trained model is effective as a feature extractor, they may contain information useful for fake audio detection.

Secondly, for the network structure, the performances of DARTS is better than LCNN. For example, we control wav2vec feature remains consistent. We can see that the EER of LCNN is higher than that of DARTS, which decreased from 3.51\% to 2.18\%. A similar conclusion is reached for wav2vec 2.0-base and wav2vec 2.0-large. This result may be explained by the fact that the advantage of darts over LCNN is that it automatically tunes the parameters of the network structure, thus eliminating the need for manual parameter setting. A surprising aspect of the data is in the result of LFCC feature. When we control LFCC feature remains consistent, we find that the EER of the DARTS is slightly higher than that of LCNN. This discrepancy could be attributed to parameters that yield optimal results on the validation set do not necessarily yield optimal results on the test set. These results indicate that DARTS whose architecture and parameters are learned automatically is effective for fake audio detection task.

Thirdly, the results are further improved when the network structure is replaced with light-DARTS. When we control for LFCC, wav2vec, wav2vec 2.0-base, and wav2vec 2.0-large separately to remain consistent, the results all show a significant decrease in EER. More specifically, using the wav2vec 2.0 large feature, we could achieve EER of 1.08\% for evaluation with the best checkpoint. A possible explanation for this might be that the MFM we used in light-DARTS not only separates the noisy signal from the informative signal, but also plays the role of feature selection between the two feature maps. Thus, it is effective for fake audio detection.

Finally, our proposed fully automated end-to-end system is effective for fake audio detection task. As shown in the last three row of Table 1, wav2vec features combined with light-DARTS improves the performance of detecting fake audio while eliminating the need for manual setting of parameters. More specifically, the wav2vec 2.0 large feature combined with light-DARTS achieve the EER of 1.08\%, which achieves best performance of single system.

\begin{figure}[t]
	\centering
	\includegraphics[height=6.2cm, width=\linewidth]{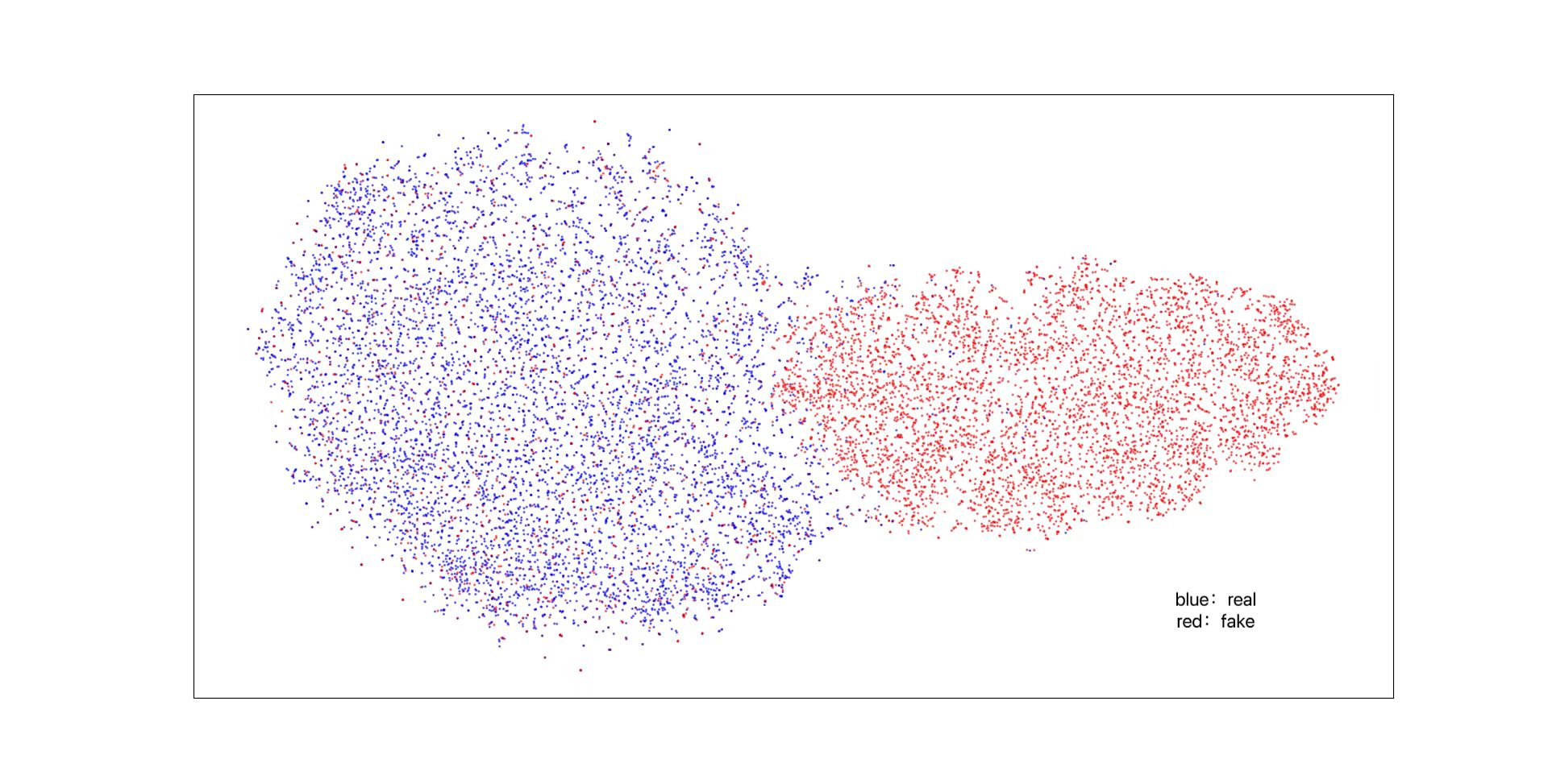}
	\caption{A t-SNE visual of LFCC-LCNN system. Blue dots represents real speech and red dots represents fake speech.}
	\Description{t-SNE for lfcc-lcnn}
\end{figure}

\begin{figure}[t]
	\centering
	\includegraphics[height=6.2cm, width=\linewidth]{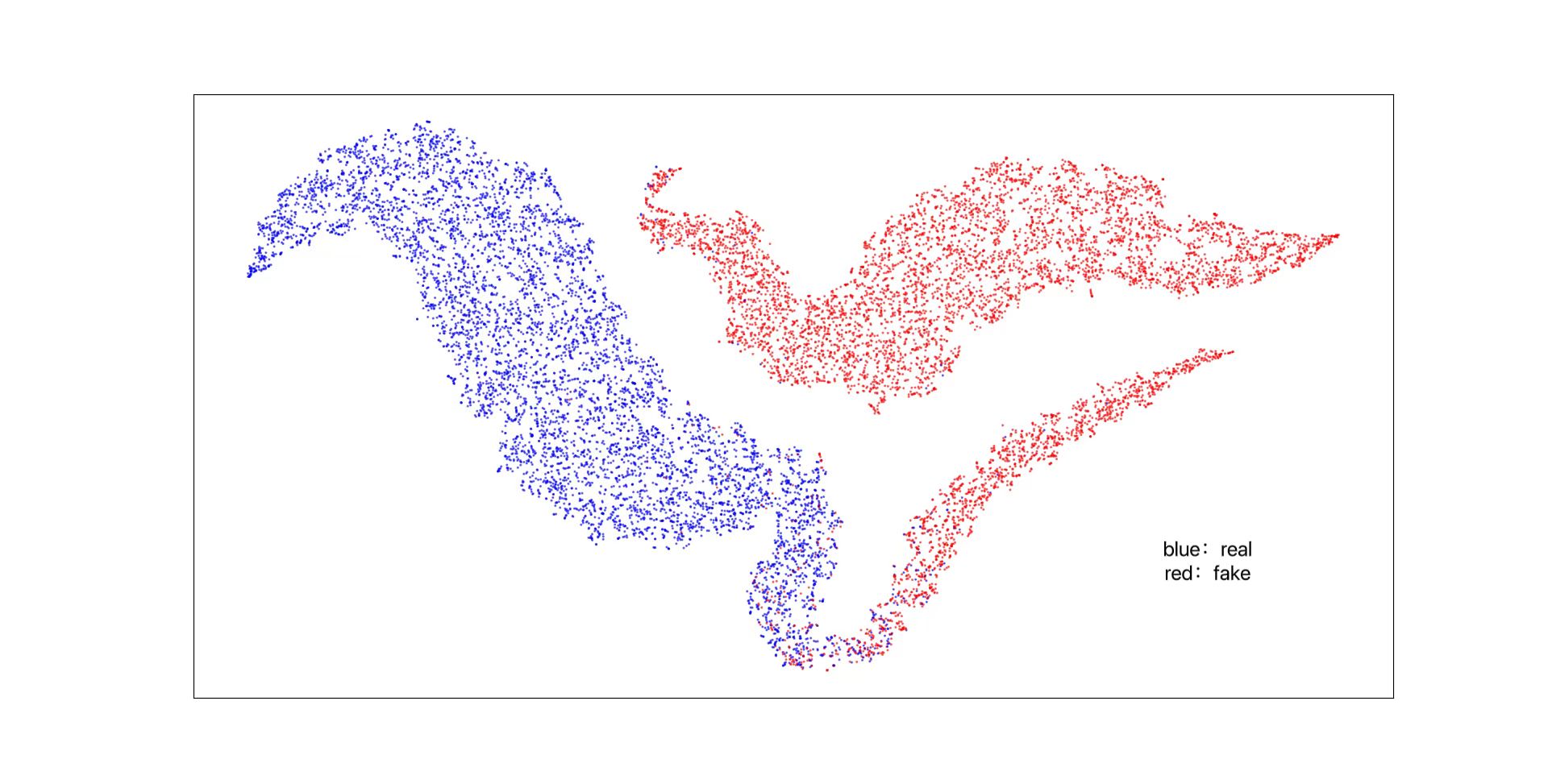}
	\caption{A t-SNE visual of wav2vec 2.0-large + light-DARTS system. Blue dots represents real speech and red dots represents fake speech.}
	\Description{t-SNE for wav2vec-light-DARTS}
\end{figure}

\subsection{Results on ASVspoof2021}
To verify the generalizability of our proposed method, we also tested it on ASVspoof2021 DF dataset. The results of the correlational analysis are presented in Table 2. The overall conclusion remains consistent with section 5.1. More specifically, the “wav2vec 2.0 large + light-DARTS” reduces the EER to 7.86\% on DF. In addition, our proposed method compares far better with ASVspoof2021 DF track first place. The above results demonstrate the generalizability of our proposed method to different datasets.

\subsection{Results on ADD 2022}
In order to further validate the effectiveness of the proposed method on Chinese dataset, we have conducted relevant experiments on ADD 2022 Track 1. Table 3 provides the experimental data on ADD 2022 Track 1. The results show that the “wav2vec 2.0-base + light-DARTS” and “wav2vec 2.0-base + light-DARTS” achieved EERs of 21.23\% and 20.11\%, respectively, on the test set, both better than the first place in the competition results. The above results show that the proposed method works on the Chinese dataset as well.

\subsection{Comparison with Other Systems}
In order to evaluate the performance of our proposed method, we also compared the proposed system with other state-of-the-art systems on the evaluation set of the ASVspoof 2019 LA database. Recent published novel systems are compared, such as sub-band models \cite{zhang122021effect}, raw waveform based models \cite{tak2021end} and frequency attention model \cite{ling2021attention}. Table 4 shows the performance comparison with other systems on the evaluation set of the ASVspoof 2019 LA database. From Table 4 we can find that our proposed fully automated end-to-end system achieves an EER of 1.08\%, outperforms the second-ranked system (EER 1.12\%) among all known systems. These results verify that the proposed method is quite effective for fake audio detection. 

\subsection{Visualization Analysis}

To visualize the effectiveness of the proposed method, we also used t-SNE \cite{van2008visualizing} to visualize LFCC-LCNN and wav2vec 2.0-large + light-DARTS, respectively. Both models are trained on the asvspoof2019 LA dataset and take the penultimate layer of the network. As shown in Figure 2 and Figure 3, we can see that the real and fake speech of LFCC-LCNN are not clearly distinguished, and there are many red dots embedded inside the blue dots. The real and fake speech of wav2vec 2.0-large + light-DARTS, on the other hand, is clearly separated, and only a little mixed at the boundary. The above visualization results are consistent with our experimental results.

\section{Conclusion}
The existing fake audio detection systems often rely on expert experience to design the acoustic features or manually design the hyperparameters of the network structure. However, artificial adjustment of the parameters can have a relatively obvious influence on the results. It is almost impossible to manually set the best set of parameters. In order to reduce the impact of manually set parameters on the stability of the fake audio detection system, this paper designs a fully automated end-to-end fake audio detection method. For features, we use wav2vec pre-trained model to obtain a high-level representation of the speech. As for classification model, we propose a light-DARTS to learn deep speech representations automatically. The results show that our proposed method achieves better detection results while not requiring manual parameter setting. The EER of our proposed method is relatively reduced by 77.26\% compared to the baseline. In the future, we will further explore the application of DARTS on fake audio detection.

\begin{acks}
This work is supported by the National Key Research \& Development Plan of China (No.2017YFC0820602), the National Natural Science Foundation of China (NSFC) (No.61831022, No.61901473, No.61771472, No.61773379 ) and Inria-CAS Joint Research Project (No.173211KYSB20190049).
\end{acks}


\bibliographystyle{unsrt}
\bibliography{acmart}
\appendix

\end{document}